\documentclass[11pt]{article}
\usepackage{appb_my,epsf,epsfig}
\begin{document}
\noindent
\thispagestyle{empty}
\begin{flushright}
{\bf DESY 99-045}\\
{\bf hep-ph/9904243}\\ 
{\bf April 1999}\\
\end{flushright}

\begin{center}
\begin{Large}
 \begingroup
   \def\thefootnote{\fnsymbol{footnote}}%
Top Quark Physics\footnote{Presented at the {\em Cracow Epiphany 
Conference on Electron-Positron Colliders}, 5-10 January 1999, Krakow, 
Poland.}\\
 \endgroup
\end{Large}
\setcounter{footnote}{0}

\vspace{0.5cm}

\begin{large}
Thomas Teubner\\[2mm]
\end{large}
Deutsches Elektronen-Synchrotron DESY, D-22603 Hamburg, Germany\\
\end{center}
\newpage

\title{Top Quark Physics\thanks{Presented at the {\em Cracow Epiphany 
Conference on Electron-Positron Colliders}, 5-10 January 1999, Krakow, 
Poland.}}
\author{Thomas Teubner
\address{Deutsches Elektronen-Synchrotron DESY, D-22607 Hamburg, Germany}}
\maketitle
\begin{abstract}
In this contribution I review the physics of top quarks at a 
future Linear Collider. Main emphasis is put on the process 
$e^+ e^- \to t\bar t$ close to threshold. Different physical observables, 
their sensitivity to the basic parameters and their theoretical prediction 
are discussed. Recent higher order calculations are shown to have a 
considerable impact on a precise determination of the top quark mass. It 
is pointed out how the use of mass definitions different from the pole 
mass scheme become important in this respect. Continuum top quark
production above threshold is discussed briefly.
\end{abstract}
\PACS{14.65.Ha, 13.90.+i, 12.38.Bx}

\section{Introduction}
Top Quark Physics will be one of the main physics cases for future collider 
physics.  Whereas the first direct discovery of top was one of the main 
successes of the proton collider at Fermilab, the precise measurement of the 
top quark mass and its couplings will remain the task of a future lepton 
collider. 

But why should we be interested in such high precision measurements in
the top sector?  
The top quark is the heaviest elementary particle observed up to now.  
Because of its very high mass $m_t \approx 175$ GeV it plays a prominent 
role for our understanding of the Standard Model (SM) and the physics 
beyond.  Already before its direct observation there was indirect evidence 
of the 
large top quark mass:  through radiative corrections $m_t$ enters 
quadratically into the $\rho$ parameter.  From precision measurements of 
the electroweak parameters $M_Z$, $M_W$, $\sin^2\theta_W$ and $G_F$ a top 
quark mass was predicted in striking agreement with the value measured at 
Fermilab.  Within the
framework of the SM the mass of the Higgs boson can be constrained from the
weak boson masses $M_W$ and $M_Z$ together with $m_t$: 
$M_H = f(M_Z, M_W, m_t)$.  As the Higgs mass enters in 
logarithmic form, stringent mass bounds can be derived only once the other 
parameters are known with high accuracy. With an absolute uncertainty of the 
top quark mass $\Delta m_t \stackrel{\scriptstyle <}{\scriptstyle \sim} 200$ 
MeV the 
Higgs mass will be extracted with an accuracy better than 17\%.  This will 
constitute one of the strongest tests of the mechanism of electroweak 
symmetry breaking at the quantum level and therefore of our understanding 
of the structure of the SM.

At the starting time of a future Linear Collider (LC) Higgs boson(s) may 
hopefully already have been discovered with the hadron machine at
Fermilab or at the LHC (assuming LEP2 is not the lucky one in the
next future).  Still, to pin down parameters precisely and to learn
which sort of physics beyond the SM 
is realized in nature, many detailed studies will be required.  With an 
expected accuracy of $\Delta m_t/m_t \approx 1\cdot 10^{-3}$ ($\Delta
m_b/m_b \simeq {\cal O}(\%)$) and the large Yukawa coupling
$\lambda_t^2 \approx 0.5$ ($\lambda_b^2 \approx 4\cdot 10^{-4}$) the
top quark will play a key role in finding the theory that gives the
link between masses and mixings and quarks and leptons.

Apart from that the large top quark mass has another important consequence: 
being much heavier than the $W$ boson the top decays predominantly into the 
$W$ and a bottom quark with the large (Born) decay rate 
\begin{equation}
\Gamma_t^{(0)} = \frac{G_F}{\sqrt{2}}\frac{m_t^3}{8\pi} \approx 1.5\ {\rm GeV}
 \gg \Lambda_{\rm QCD}\,.
\end{equation}
Therefore top is the only quark that lives too short to hadronize.  The large 
width $\Gamma_t$ serves as a welcome cut-off of non-perturbative 
effects \cite{K} and the top quark behaves like a {\em free} quark.  In 
this way top quark physics is an ideal test-laboratory for QCD at high 
scales, where predictions within perturbation theory are reliable.

Having these goals in mind a future $e^+ e^-$ Linear Collider (see
e.g.~\cite{Designrep, PRep}) will be the ideal machine to study the
top quark in detail. (The same will be true for a $\mu^+ \mu^-$
collider, once technologically feasible.)  The clean environment and
generally small backgrounds make it complementary to hadron machines,
where higher energies can be achieved more easily.
In addition the collision of point-like, colourless leptons 
guarantees very good control of the systematic uncertainties.  Operation with 
highly polarized electrons (and to a smaller extent also positrons) 
is realizable and will open new possibilities.  Another option is 
the use of Compton back-scattered 
photons of intense lasers from the electron and positron bunches, allowing 
for operation of the $e^+ e^-$ collider in the $\gamma \gamma$ (or 
$e \gamma$) mode.  These modes can be very useful for certain studies of the 
Higgs sector and other areas of electroweak physics, but will be less 
important for top quark physics.  Therefore the following discussion will be 
limited to $e^+ e^-$ collisions.\footnote{Reader interested in the physics of 
$\gamma \gamma$ collisions are referred to \cite{Designrep} (and
references therein) for a general discussion and to \cite{gammagamma}
especially for $\gamma \gamma \to t \bar t$ at threshold.}

The article is organized as follows:  In Section 2 the scenario of top quark 
pair production at threshold is described in some detail.  I discuss the 
important parameters, accessible observables and their sensitivity, and the 
corresponding theoretical predictions.  Recent higher order calculations are 
reviewed.  It will be shown how the large theoretical uncertainties in the 
shape of the cross section near threshold can be avoided by using a mass 
definition different from the pole mass scheme.  In Section 3 a brief 
discussion of some important issues in top quark production above threshold 
is given.  Section 4 contains the conclusions.  For a
comprehensive review of top quark physics (including top at hadron colliders)
see also \cite{Kslacrep}.  Clearly the rich field of top quark physics
cannot be completely covered in this contribution, which is somewhat
biased towards $t\bar t$ at threshold.  This is also partly due to
the author's experience. I would like to apologize to those who miss
important information or feel own contributions to top physics not
covered properly or not mentioned at all.

\section{The $t\bar t$ Threshold}
\subsection{What's so special about the top threshold?}
Close to the nominal production threshold $\sqrt{s} = 2 m_t$ top and antitop 
are produced with non-relativistic velocities 
$v = \sqrt{1 - 4 m_t^2/s} \ll 1$.  The exchange of (multiple,
ladder-like) Coulombic gluons leads 
to a strong attractive interaction, proportional to $(\alpha_s/v)^n$.  These 
terms are not suppressed and the usual expansion in $\alpha_s$ breaks down.  
Summation leads to the well known Coulomb enhancement factor at threshold, 
giving a smooth transition to the regime of bound state formation below 
threshold, which cannot be described using ordinary perturbation theory.  
In principle we would expect a picture like this with ``Toponium'' resonances 
similar to the case of bottom quarks which form the $\Upsilon(nS)$
mesons at threshold.  
However, in the case of top quarks, the rapid decay makes a formation of real 
$t\bar t$ bound states impossible.  The width of the $t\bar t$ system is 
saturated by the decay of its constituents: $\Gamma_{t-\bar t} \approx 
2 \Gamma_t \approx 3$ GeV.  This is much larger than the 
expected level spacing and leads to a smearing of any sharp resonance 
structure, leaving only a remnant of the $1S$ peak visible in the excitation 
curve.  Therefore there will be nothing like $t\bar t$-spectroscopy to study 
at the top threshold.  Nevertheless the short life-time of the top quarks 
also has a remarkable advantage:  non-perturbative effects, 
hadronization and real (soft) gluon emission are suppressed by $\Gamma_t$, 
$m_t$.\footnote{For studies concerning the effects of real gluon
  emission see also Ref.~\cite{Orr}.} 
Therefore, in contrast to the bottom (let alone the charm) quark 
sector, top quark production becomes calculable in perturbative QCD 
\cite{FK}.  
$t\bar t$ is, from the theoretical point of view, much ``cleaner'' than 
$c\bar c$ and $b\bar b$ and will allow for more detailed tests of the 
underlying theory and a more precise determination of the basic parameters 
$m_t$, $\alpha_s$ (and $\Gamma_t$).  In this sense $t\bar t$ at threshold 
is a unique system, which deserves to be studied in detail at a 
future $e^+ e^-$ collider.

\subsection{Parameters to be determined}
$\bullet$ As mentioned already above the main goal will be a precise 
measurement of the top quark mass.  Current analyses from CDF and D0 
at the Tevatron at 
Fermilab determine $m_t$ by reconstructing the mass event by event.  
Current values are
\begin{eqnarray}
m_t^{\rm pole} &=& 176.0 \pm 6.5\ {\rm GeV}\qquad ({\rm CDF}\ \cite{CDF})\,, 
\nonumber\\
m_t^{\rm pole} &=& 172.1 \pm 7.1\ {\rm GeV}\qquad ({\rm D0}\ \cite{D0})\,.
\end{eqnarray}
The Run II at the Tevatron is expected to improve the accuracy down to maybe
$\Delta m_t = 2$ GeV.  It looks impossible to reach a higher 
accuracy at hadron colliders.  In contrast, with a threshold scan of the 
cross section at a future $e^+ e^-$ Linear Collider one will be able to reach 
$\Delta m_t = 200$ MeV or even better \cite{PRep, MM, FMY}.  
High luminosity will 
allow for very small statistical errors so that the accuracy will be limited 
mainly by systematic errors and theoretical uncertainties.  

$\bullet$ The strong coupling $\alpha_s$ governs the interaction of $t$ and 
$\bar t$.  It enters the Coulombic potential $V(r) = -C_F\,\alpha_s/r$ 
which dominates close to threshold, as well as other corrections which get 
important at higher orders of perturbation theory (see below).  $\alpha_s$ 
may either be taken as an input (with some error) measured independently at 
other experiments or, alternatively, can be determined simultaneously with 
$m_t$ in a combined fit.

$\bullet$ The (free) top quark width $\Gamma_t$ leads to the smearing of the 
resonances and strongly influences the shape of the cross section at 
threshold.  As will be discussed below, $\Gamma_t$ can be measured with 
good precision near threshold either in the $t\bar t$ production process 
or by help of observables specific to the decay.\footnote{For a
  detailed discussion of top quark decays see also \cite{Jnpbproc}.}
In the framework of the 
SM $\Gamma_t$ can be predicted reliably:  the first order $\alpha_s$ \cite{JK}
and electroweak \cite{DSao} corrections are known for some time (see
also \cite{JK2}), and recently even corrections of order $\alpha_s^2$ 
became available \cite{CM}.  The ${\cal O}(\alpha_s)$ corrections
lower the Born result by about 10\%, whereas ${\cal O}(\alpha_s^2)$ and 
electroweak contributions effectively cancel each other, with
corrections of about $-2\%$ and $+2\%$, respectively.  In extensions of the
SM the top quark decay rate can be significantly different from the SM
value:  new channels like the decay in a charged Higgs ($t \to b H^+$)
in supersymmetric theories will lead to an increase of $\Gamma_t$.  In
models with a forth generation the Cabibbo-Kobayashi-Maskawa (CKM) 
quark-mixing-matrix element $V_{tb}$ will be smaller than the SM value 
$V_{tb}^{({\rm SM})} \simeq 1$ and lead to a suppression of 
$\Gamma_t^{({\rm SM})}$.

$\bullet$ The electroweak couplings of the top quark enter both in
production and decay.  Especially in angular distributions (of the decay
products) and in observables sensitive to the polarization of the top
quarks deviations from the SM may be found.  In principle even the
influence of the Higgs on the $t\bar t$ production vertex should be
visible \cite{HJK}.  Unfortunately, for the currently allowed range of
Higgs-masses, effects due to (heavy) Higgs exchange mainly result in a
``hard'' vertex correction which changes the overall normalization of
the cross section.  As will be discussed below, contributions of this
sort are in competition with uncertainties from other higher order 
corrections and therefore difficult to disentangle at the $t\bar t$ 
threshold.\\

Therefore, to determine the parameters with high precision and to
eventually become sensitive to new physics, a thorough understanding
of the SM physics, in particular the QCD dynamics, is mandatory.

\subsection{Theory's tools to make predictions}
\label{theorytools}
How to predict the cross section close to threshold? In principle one
could write the cross section as a sum over many overlapping
resonances \cite{Kwong}:
\begin{equation}
\sigma(e^+ e^- \to t\bar t\,) \sim -{\rm Im}\, \sum_n 
\frac{|\psi_n(r=0)|^2}{E-E_n+i \Gamma_t}\,,
\end{equation}
where $\psi_n$ are the wave functions of the $nS$ states with the
corresponding Eigenenergies $E_n$. (Close to threshold $S$ wave
production is dominating with the contributions from $P$ waves being
suppressed by two powers of the velocity $v$. With $v \approx \alpha_s$
these contributions have to be considered only at
next-to-next-to-leading order.)  However, this explicit summation is not
very convenient, as the sum does not converge fast, especially for positive
energies $E = \sqrt{s} - 2 m_t$.  As shown by Fadin and Khoze
\cite{FK}, the problem can be solved within the formalism of 
non-relativistic Green functions:
\begin{equation}
\sigma(e^+ e^- \to t\bar t\,) \sim -{\rm Im}\, G(r=0, E+i\Gamma_t)\,. 
\end{equation}
The Green function $G$ is the solution of the Schr\"odinger equation
\begin{equation}
\left[ \left( -\frac{\vec\nabla^2}{m_t} + V\left(\vec r\,\right) \right) - 
\left(E+i\Gamma_t\right) \right] G\left(\vec r, E+i\Gamma_t\right) = 
\delta^{(3)}\left(\vec r\,\right)
\end{equation}
or, equivalently, the Lippmann-Schwinger equation in momentum space
\begin{equation}
\tilde G\left(\vec p, E+i\Gamma_t\right) = \tilde G_0 + \tilde G_0 
\int \frac{{\rm d}^3 q}{(2\pi)^3} \tilde V\left(\vec p - \vec q\,\right) 
\tilde G\left(\vec q, E+i\Gamma_t\right) \,,
\end{equation}
where $\tilde G_0 \equiv \left(E+i\Gamma_t-p^2/m_t\right)^{-1}$ is
the free Green function.  At leading and next-to-leading order the
continuation of the energy in the complex plane $E+i\Gamma_t$ is all
that is needed to take care of the finite decay width of the top
quarks.  These equations can be solved numerically using a realistic
QCD potential $V(r) = -C_F \alpha_s(r)/r$ or $\tilde V\left(q^2\right)
= - 4 \pi C_F \alpha_s(q^2)/q^2$ to give the total cross 
section \cite{StrasslerPeskin, Sumino, JKT}  
\begin{equation}
\sigma\left(e^+ e^- \to \gamma^* \to t\bar t\,\right) = 
\frac{32\,\pi^2\,\alpha^2}{3\,m_t^2\,s}\,
{\rm Im}\, G\left(r=0, E+i\Gamma_t\right)\,.
\label{eqsigmatot}
\end{equation}
The top quark momentum
distribution (differential with respect to the modulus of the top
quark three momentum $p$), which reflects the Fermi motion in the
would-be bound state and the instability of the top quarks, is obtained by
\begin{equation}
\frac{{\rm d}\sigma(p, E+i\Gamma_t)}{{\rm d}p} = 
\frac{16\,\alpha^2}{3\,s\,m_t^2}\,
 \Gamma_t \, p^2 \left| \tilde G\left(p, E+i\Gamma_t\right)\right|^2\,.
\label{eqsigmadif}
\end{equation}
Eqs.~(\ref{eqsigmatot}, \ref{eqsigmadif}) are correct at leading order
in $\alpha_s$, $v$.  At next-to-leading order (NLO) various new effects have
to be taken into account.  Apart from the well known ${\cal
  O}(\alpha_s)$ corrections to the static QCD potential \cite{FandB} the
exchange of ``hard'' gluons results in 
the vertex correction factor $\left( 1 - 16 \alpha_s/(3\pi)\right)$ in
the (total and differential) cross section \cite{Barbierietal}.
Interference of the production through a virtual photon and a virtual
$Z$ boson leads to the interference of the vector current induced $S$
wave with the axial-vector current induced $P$ wave contributions.  
This $S$-$P$ wave interference is suppressed by order $v$ and drops out
in the total cross section after the angular ($\cos\theta$)
integration.  However, it contributes to the differential rate and
will be measured in observables like the forward-backward asymmetry 
${\cal A}_{\rm FB}$ \cite{Sumino2, HJKT}.  In addition, at order
$\alpha_s$, there are final state corrections coming from gluon
exchange between the produced $t$ and $\bar t$ and their strong
interacting decay products $b$ and $\bar b$.  The final state
interactions in the $t b$ and $\bar t \bar b$ systems factorize and
are easily taken into account by using the (order $\alpha_s$) corrected
free top quark width $\Gamma_t$, with no other corrections at ${\cal
  O}(\alpha_s)$ \cite{JT, MK}.  However, the ``crosstalk'' between
$t$ $\bar b$, $\bar t$ $b$ and $b$ $\bar b$ leads to
non-factorizable corrections which have to be considered in
addition.\footnote{In principle hadronically decaying $W$ bosons also
  take part in these final state interactions.}
These corrections are suppressed in the total cross section \cite{FKM,
  MY}, but contribute in differential distributions and hence in 
${\cal A}_{\rm FB}$.\footnote{See also Ref.~\cite{khoze} and references
  therein for a discussion of the possible impact of colour
  reconnection effects on the top quark mass determination.}  
Results obtained in the framework of the non-relativistic Green
function approach are available at ${\cal O}(\alpha_s)$, see
\cite{HJKP, PS}.  
At the same accuracy polarization of the produced
$t$ and $\bar t$, depending on the polarization of the $e^+$ and $e^-$
beams, has been studied in \cite{HJKT, HJKP, PS}.  
Therefore, at order $\left(\alpha_s,\,v\right)$, theoretical predictions are
available for a variety of observables at the top quark threshold, as
will be discussed in the next paragraph.

Electroweak corrections to the $t \bar t$ production vertex have been
calculated for the threshold region \cite{GK} as well as for general
energies \cite{BHMandBH} in the SM and even in the Minimal
Supersymmetric SM (MSSM), see \cite{HS}.  

\subsection{Observables and their sensitivity}
\begin{figure}[htb]
\vspace{-0.5cm}
\begin{center}
\leavevmode
\epsfxsize=10.0cm
\epsffile[105 275 475 555]{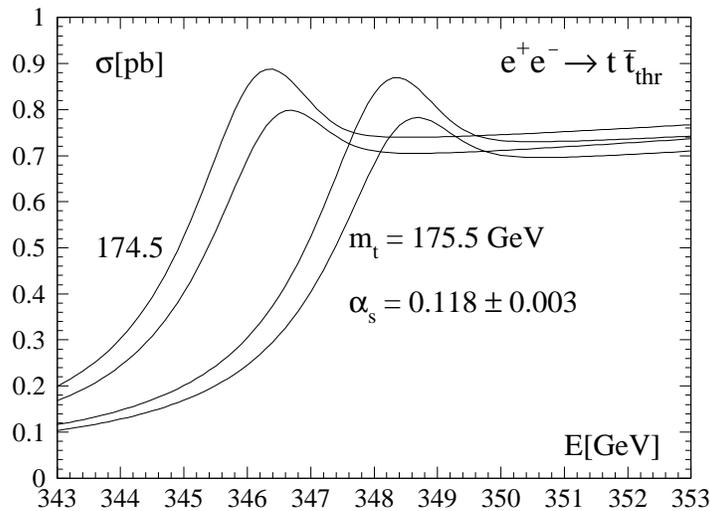}
\end{center}
\vspace{-0.5cm}
\caption[]{\label{fig1} Total cross section $\sigma(e^+ e^- \to t\bar
  t\,)$ (in pb) as a function of the total centre of mass energy for two
  different values of $m_t$ and $\alpha_s$.  The upper curves
  correspond to $\alpha_s(M_Z) = 0.121$, the lower ones to
  $\alpha_s(M_Z) = 0.115$. (Figure taken from \cite{PRep}.)}
\end{figure}
$\bullet$ The (from the theoretical as well as from the experimatal
point of view) cleanest observable is the total cross section
$\sigma_{\rm tot}$.
Depending on the decays of the $W^+$ and $W^-$ from the $t$ and the
$\bar t$ quarks, $t\bar t$ decays into six jets (46\%), four jets + $l +
\nu_l$ (44\%) or two jets + $l\,l'\,\nu\,\nu'$ (10\%) (60, 35
and 5\%, respectively, if $l = e,\,\mu$ only and $\tau$-leptons are
excluded).  The main backgrounds are from $e^+ e^- \to W^+ W^-$, $Z^0
Z^0$ and $f \bar f\,$(plus gluons and photons).  These processes are
well under control as distinguishable from the signal (e.g.\ by higher
Thrust or less jets) and constitute no big problem for the experimental
analysis.  The total cross section is mainly sensitive to $m_t$ and
$\alpha_s$.  Fig.~\ref{fig1} shows the cross section for two different
values of the top quark mass and two values of the strong coupling,
plotted over the total centre of mass energy.  Note the correlation
between $m_t$ and $\alpha_s$:  higher top-masses lead to a shift of
the remainder of the $1S$ peak to larger energies.  In a similar way
an increase of $\alpha_s$ is equivalent to a stronger potential (a
larger negative binding energy) and hence lowers the peak position.
I will come back to this point later.  In practice, the shape of the
cross section will not look as pronounced as in Fig.~\ref{fig1}.
\begin{figure}[htb]
\begin{center}
\epsfig{file=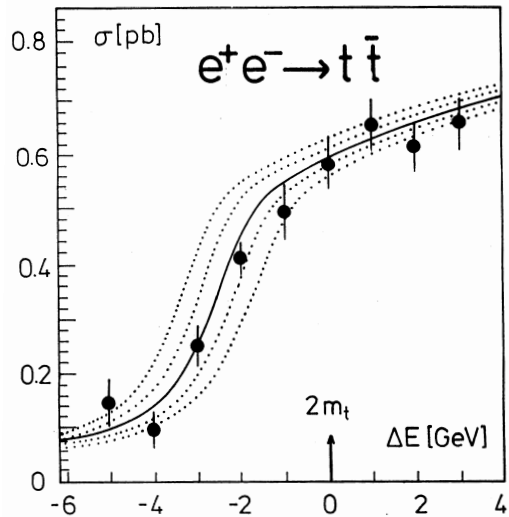,width=7cm,angle=89.5}
\end{center}
\vspace{-.2cm}
\caption[]{\label{fig2} Total cross section in the threshold region
  including initial-state and beam-strahlung.  The errors of the data
  points correspond to an integrated luminosity of $\int{\cal L} = 50\
  {\rm fb}^{-1}$ in total.  The dotted curves
  indicate shifts of the top mass by 200 and 400 MeV.  (Figure taken
  from \cite{PRep}.)}
\end{figure}
Initial state radiation (of photons from the $e^+$ and $e^-$ beams) as
well as the beamstrahlung-effects from the interaction of the $e^+$
and $e^-$ bunches lead to a distortion of the original shape.
Fig.~\ref{fig2} displays how the total cross section is expected
to look under realistic conditions.  The dots in the plot are
Monte-Carlo generated ``data points'' of a typical planned threshold
scan.

In addition $\sigma_{\rm tot}$ also depends on the Higgs mass $M_H$
and the top quark width $\Gamma_t$.  
As mentioned already above, the Higgs mainly influences the
normalization of the cross section which will probably not allow for a
high sensitivity to $M_H$ once other uncertainties are taken into account.
$\Gamma_t$, on the other hand, influences the shape:  the smaller the
width the more pronounced the peak.  This will be used together with
the sensitivity of other observables to measure $\Gamma_t$.\\

\begin{figure}[htb]
\begin{center}
\leavevmode
\epsfxsize=10.0cm
\epsffile[105 275 475 555]{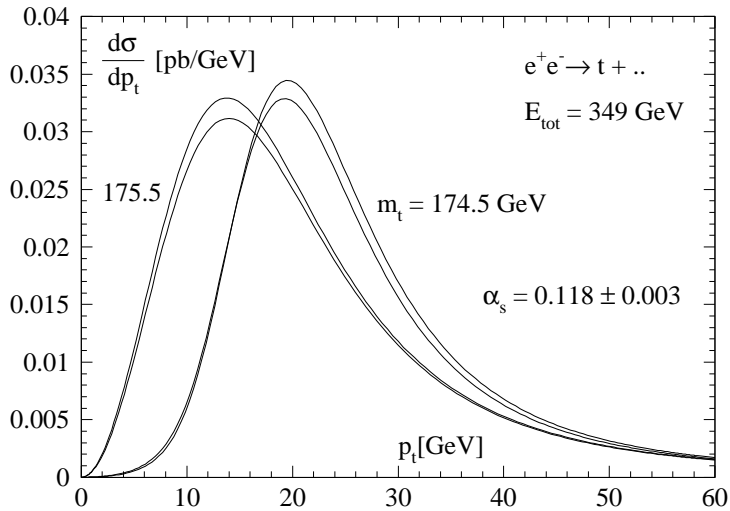}
\end{center}
\vspace{-0.5cm}
\caption[]{\label{fig3} The differential cross section ${\rm
    d}\sigma/{\rm d}p$ as a function of the top quark momentum $p$ for
    a fixed value of the centre of mass energy (349 GeV).  $m_t$ and
    $\alpha_s$ are chosen as indicated.  (Figure taken from
    \cite{PRep}.)}
\end{figure}

$\bullet$ Another observable is the momentum distribution 
${\rm d}\sigma/{\rm d}p$, obtained from the reconstruction of the
three momentum of the top (and antitop) quark.  With the possible high
statistics at a future Linear Collider the distribution can be well
measured.  As shown in Fig.~\ref{fig3}, the peak position strongly
depends on $m_t$ but less on the QCD coupling:  for higher values of
$m_t$ the distribution is peaked at much lower momenta, whereas
the coupling strength mainly changes the normalization.  
Therefore a
measurement of the momentum distribution can help to disentangle the
strong correlation of $m_t$ and $\alpha_s$ in the total cross section
(see \cite{MM}).  There is also a less pronounced dependence on
$\Gamma_t$.\\

$\bullet$ As mentioned above, $S$-$P$ wave interference leads to a 
nontrivial $\cos\theta$ ($\theta$ being the angle between the $e-$
beam and the $t$ direction) dependence of the cross section.  The
resulting forward-backward asymmetry
\begin{equation}
{\cal A}_{\rm FB} = \frac{1}{\sigma_{\rm tot}} \left[
\int_0^1 {\rm d}\cos\theta - \int_{-1}^0 {\rm d}\cos\theta \right]
\frac{{\rm d}\sigma}{{\rm d}\cos\theta}
\label{defafb}
\end{equation}
shows a considerable dependence on $\Gamma_t$ and $\alpha_s$, but is not
very sensitive to $m_t$.  
\begin{figure}[htb]
\begin{center}
\leavevmode
\epsfxsize=9.0cm
\epsffile[90 150 470 655]{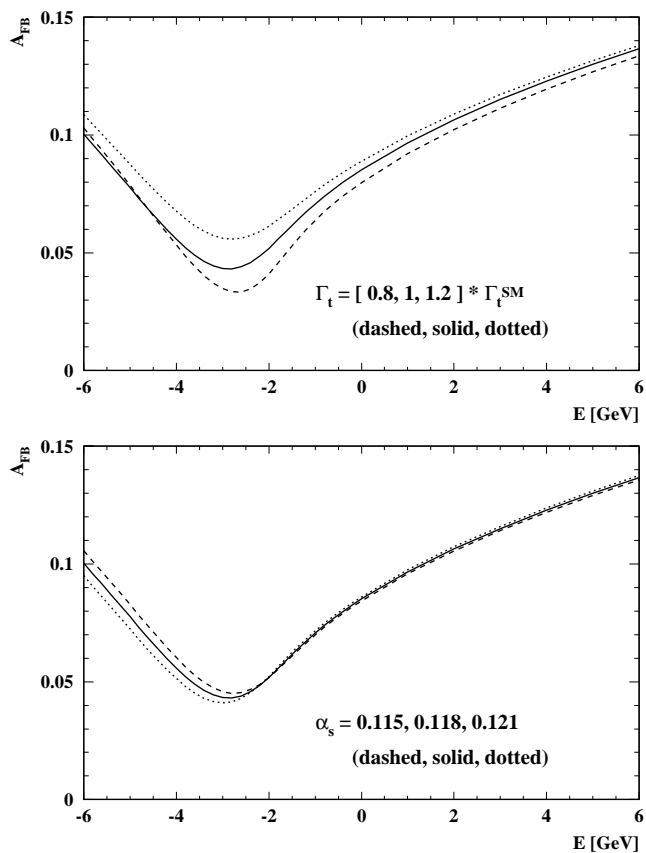}
\end{center}
\vspace{-0.5cm}
\caption[]{\label{fig4} Forward-backward asymmetry ${\cal A}_{\rm FB}$
  as a function of $E = \sqrt{s} - 2 m_t$ for three different values
  of the top quark width and the strong coupling.  Upper plot:
  variation of $\Gamma_t$ by $\pm 20\%$ around the SM value 
  $\Gamma_t^{\rm SM} = 1.43$ GeV and 
  $\alpha_s(M_Z) = 0.118$.  Lower plot:
  $\alpha_s(M_Z) = 0.115, 0.118, 0.121$ and $\Gamma_t = 1.43$ GeV. 
  ($m_t = 175$ GeV.)}
\end{figure}
\begin{figure}[htb]
\vspace{-0.5cm}
\begin{center}
\leavevmode
\epsfxsize=8.0cm
\epsffile{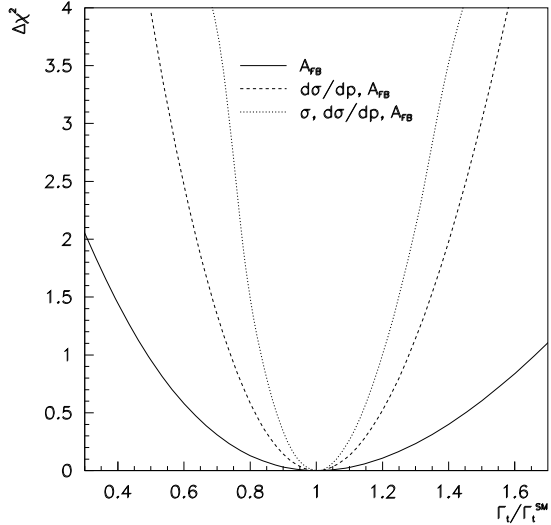}
\end{center}
\vspace{-0.5cm}
\caption[]{\label{fig5} Increase of the $\chi^2$ of the fit as a
function of the top quark width using the forward-backward asymmetry
${\cal A}_{\rm FB}$ (solid line), adding the top quark momentum
distribution (dashed line) and the total cross section (dotted line).
(Figure taken from \cite{MM}.)}
\end{figure}
In Fig.~\ref{fig4} ${\cal A}_{\rm FB}$ is
plotted as a function of $\sqrt{s}$ for three different choices of
$\Gamma_t$ and $\alpha_s$.  With 
increasing width the overlap of $S$ and $P$ waves
becomes bigger and hence the asymmetry is enhanced.  Together with the
total and differential cross section
the measurement of ${\cal A}_{\rm FB}$ can be used to determine
$\Gamma_t$ by a fit.
The sensitivity of such a fit to the different observables is
demonstrated in Fig.~\ref{fig5}.\\

Please note that the figures for the cross section and the asymmetry
do not contain the (nonfactorizable) ${\cal O}(\alpha_s)$ rescattering
corrections discussed in Section \ref{theorytools}.  They are absent
in the total cross section but slightly change ${\rm d}\sigma/{\rm
  d}p$ and ${\cal A}_{\rm FB}$, see \cite{HJKP, PS} for a detailed
discussion.\\

$\bullet$ Top Quark Polarization:
Near threshold $S$ wave production dominates ($\vec L = 0$) and the
total spin consists of the spins of the top and antitop quarks, $\vec
J_{\gamma^*,\,Z^*} = \vec S_t + \vec S_{\bar t}$.  In leading order
the top spin is aligned with the $e^+ e^-$ beam direction.  Even without
polarization of the initial $e^+$ and $e^-$ beams, the top quarks are 
produced with $-40\%$ (longitudinal) polarization.  For a realistic
(longitudinal) $e^-$ polarization of $P_{e^-} = +80\%$ ($-80\%$) and an
unpolarized $e^+$ beam ($P_{e^+} = 0$) the
top polarization amounts to $+60\%$ ($-90\%$).  This picture is changed
only slightly due to $S$-$P$ wave interference effects of ${\cal O}(v)$
and rescattering effects of ${\cal O}(\alpha_s)$, which lead to top
polarizations perpendicular to the beam direction (transverse) and
normal to the production plane.  Normal polarization could also be
induced by time reversal odd components of the $\gamma t \bar t$- or
$Z t \bar t$-couplings, e.g.\ by an electric dipole moment, signalling
physics beyond the SM.  

The influence of the bound state dynamics near
threshold was calculated in the Green function formalism, including
the polarization of the initial beams, the $S$-$P$ wave interference
contributions and the ${\cal O}(\alpha_s)$ 
rescattering effects \cite{HJKT, HJKP, PS}.  Neglecting contributions
due to rescattering, the three polarizations can be written as
\begin{eqnarray}
\left| \vec S_{\|} \right| &=& C^0_{\|} + C^1_{\|}\,\varphi_{\rm R}(p, E)
\,\cos\theta\,,\nonumber\\
\left| \vec S_{\bot} \right| &=& C_{\bot}\,\varphi_{\rm R}(p, E)\,
\sin\theta\,,\nonumber\\
\left| \vec S_{\rm N} \right| &=& C_{\rm N}\,\varphi_{\rm I}(p, E)\,
\sin\theta\,.
\end{eqnarray}
The functions $\varphi_{\rm R, I}$ contain all information about the
threshold dynamics, whereas the coefficients $C$ are dependent on the
electroweak couplings and the $e^+ e^-$ polarization (see e.g.\
Ref.~\cite{HJKP} for complete formulae).  
\begin{figure}[htb]
\begin{center}
\leavevmode
\epsfxsize=12.0cm
\epsffile[30 285 540 530]{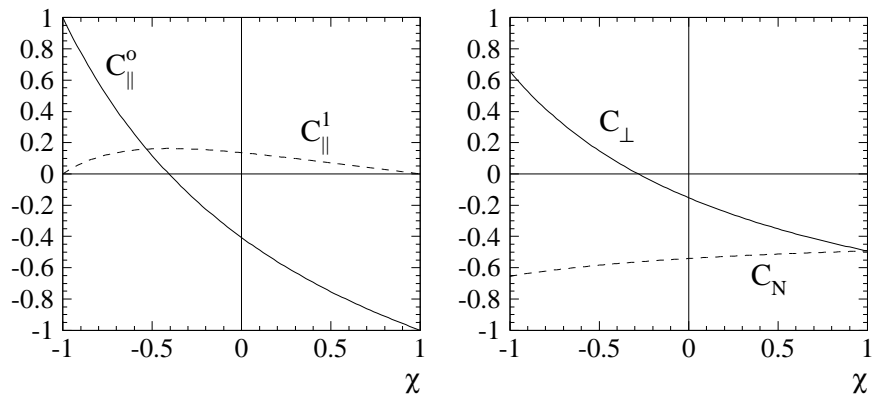}
\end{center}
\vspace{-0.5cm}
\caption[]{\label{fig6} Coefficients $C$ as functions of the
  polarization $\chi$ as described in the text, for
  $\sqrt{s} = 180$ GeV and $\sin^2\theta_W = 0.2317$.}
\end{figure}
Fig.~\ref{fig6} shows the coefficients $C^0_{\|}$, $C^0_{\bot}$,
$C_{\bot}$ and $C_{\rm N}$ as functions of the effective polarization 
$\chi = (P_{e^+} - P_{e^-})/(1 - P_{e^+}P_{e^-})$.  From
Fig.~\ref{fig6} it becomes clear that by choosing the appropriate
longitudinal polarization of the $e^-$ beam one can tune the normal
polarization of the top quarks $\vec S_{\rm N}$ to dominate.  The
functions $\varphi_{\rm R, I}(p, E)$ are displayed in Fig.~\ref{fig7} for
four different energies $E$ around the threshold.  
\begin{figure}[htb]
\begin{center}
\leavevmode
\epsfxsize=11.0cm
\epsffile[30 170 540 630]{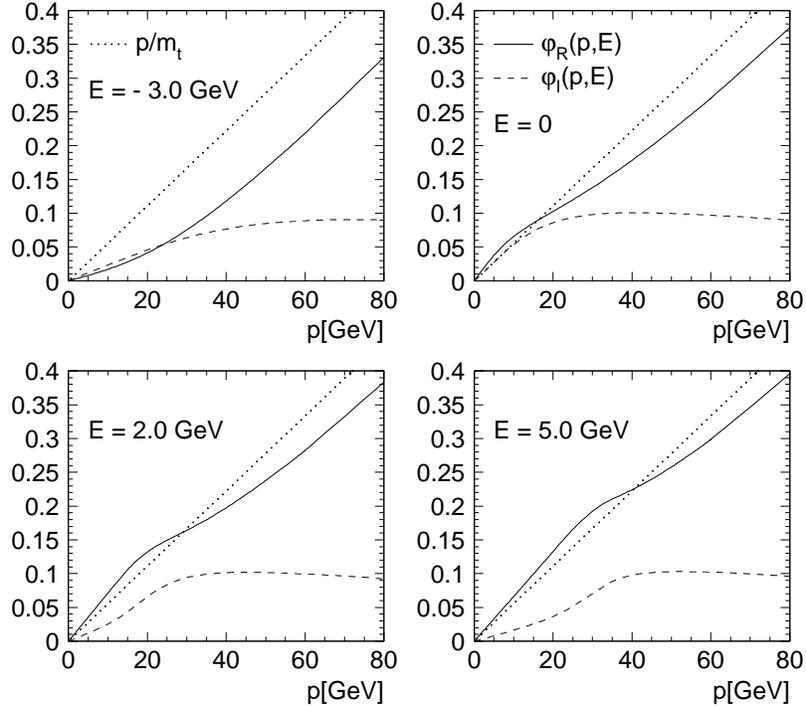}
\end{center}
\vspace{-0.5cm}
\caption[]{\label{fig7} Functions $\varphi_{\rm R}(p, E)$ (solid curves)
  and $\varphi_{\rm I}(p, E)$ (dashed) for four
  different energies close to threshold ($m_t = 180$ GeV, $\alpha_s =
  0.125$).  The dotted lines show the free particle result $\varphi_{\rm
  R} = p/m_t$. (Figure taken from \cite{HJKP}.)}
\end{figure}
Also shown is the result for free quarks, $\varphi_{\rm R} = p/m_t$.  

The normal polarization depends basically on the parameters
$\Gamma_t$, $\alpha_s$ and is relatively stable against rescattering
corrections.  The $\alpha_s$ dependence can be understood from the case of
stable quarks
and a pure Coulomb potential, where the analytical solution exists
\cite{FKK}: $\varphi_{\rm I} \to \frac{2}{3}\,\alpha_s$.  In contrast,
the subleading (angular dependent) part of the longitudinal
polarization and the transverse polarization both are (strongly)
changed by rescattering corrections, but vanish after angular
integration.  For a detailed discussion of the rescattering corrections
and the construction of inclusive and exclusive observables which are
sensitive to the top quark polarization, see \cite{HJKP, PS, P}.  
Let me just note here that the rescattering corrections
destroy the factorization of the production and decay of
the polarized top quarks.  Nevertheless, observables can be
constructed which depend neither on the subtleties of the $t\bar t$
production process nor on rescattering corrections, but only on the
decay of free polarized quarks, even in the presence of anomalous
top-decay vertices (see \cite{PS, SandJS}).\\

\begin{figure}[htb]
\begin{center}
\vspace{-0.5cm}
\leavevmode
\epsfxsize=9.5cm
\epsffile[100 120 460 700]{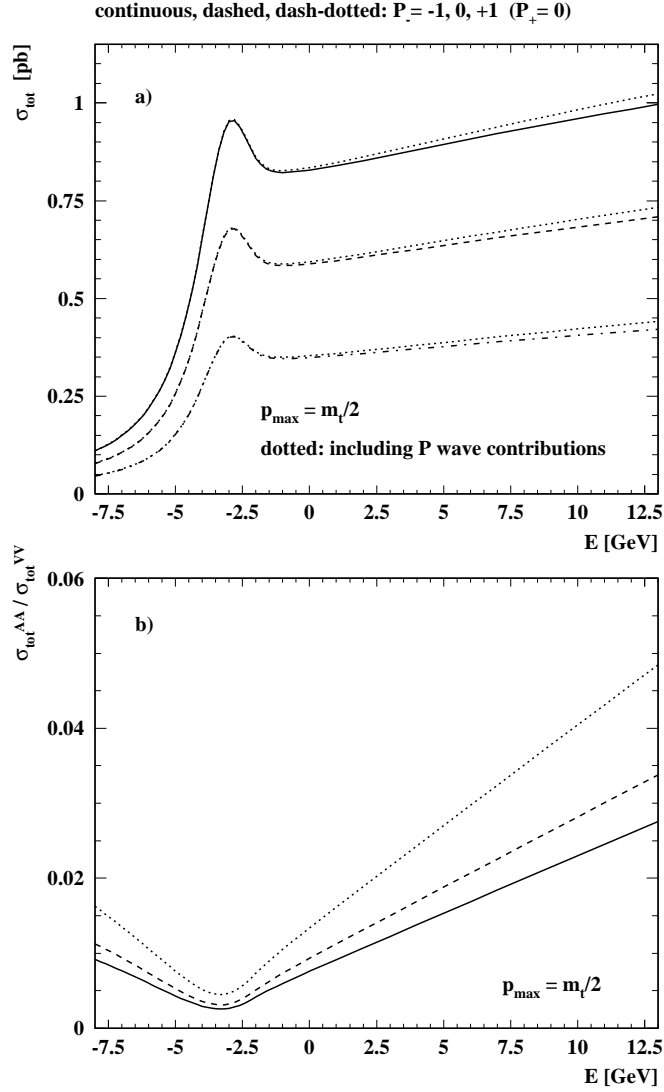}
\end{center}
\vspace{-0.5cm}
\caption[]{\label{fig8} a) The total cross 
section $\sigma(e^+ e^- \to t\bar t\,)$ 
as a function of $E$ for three different choices of the $e^-$ polarization: 
the continuous, dashed and dash-dotted lines correspond to $P_- = -1$, $0$ 
and $1$, respectively, where only $S$ wave production is taken into account. 
The dotted lines show the corresponding total cross sections including 
the $P$ wave contributions. b) Ratio of the $P$ to the $S$ wave contribution 
$\sigma_{\rm tot}^{\rm AA}/\sigma_{\rm tot}^{\rm VV}$ for the 
three different $e^-$ polarizations. (Figure taken from \cite{KT}.)}
\end{figure}

$\bullet$ Axial contributions to the angular integrated cross section:
$P$ wave contributions arise not only at ${\cal O}(v)$ due to $S$-$P$
wave interference but also as $P^2$-terms at next-to-next-to-leading
order (NNLO).  These contributions are suppressed by $v^2$ close to
threshold.  Still, they contribute at the percent level and have to be
taken into account at the NNLO-accuracy discussed below.  In addition
these axial current induced corrections are an independent
observable and strongly depend on the polarization of the $e^+ e^-$
beams.  Numerical results for the total and differential cross
section were obtained recently within the formalism of
non-relativistic Green functions \cite{KT}.  Fig.~\ref{fig8} shows the
total cross section as a function of the energy with and without these
contributions and their size relative to the pure $S$ wave result for
three different values of the $e^-$ polarization.  
A cut-off $p_{\rm max} = m_t/2$ has been
applied to cure the divergence of the integrated $P$ wave Green
function coming from the large momentum region, where the
non-relativistic approximation breaks down.

\subsection{Large next-to-next-to-leading order corrections}
In view of the size of the NLO corrections one may ask how
accurate the theoretical predictions are.  To answer this question
within perturbation theory convincingly one has to go to the next
order, in our case to the NNLO.  The first step in this direction was
done by M.~Peter who calculated the ${\cal O}(\alpha_s^2)$ corrections
to the static potential \cite{PeterSchroeder}.  They turned out to be
sizeable and, furthermore, indicate limitations of the 
accuracy achievable due to the asymptoticness of the perturbative series.
As was studied in \cite{JKPST}, the series for the effective coupling
in the Coulomb potential behaves differently in the position and in
the momentum space.  Although potentials formally may differ only in
N$^3$LO, the resulting theoretical uncertainty of the total cross
section in the $1S$ peak region is estimated to be of the order 6\%
\cite{JKPST}.  

Recently results of the complete NNLO relativistic
corrections\footnote{Here NNLO means corrections of the order ${\cal
    O}(\alpha_s^2,\,\alpha_s v,\,v^2)$ relative to the Born result
  which contains the resummation of the leading $(\alpha_s/v)^n$
  terms.} to $t \bar t$ production near threshold became available
\cite{HT2, MYel, Yakovlev, BSS}.  
The results are in fair agreement
and modify the NLO prediction considerably.  In the following I will
briefly describe the calculation and results.\\

\begin{figure}[htb]
\begin{center}
\vspace{0.5cm}
\leavevmode
\epsfxsize=3.5cm
\epsffile[220 420 420 550]{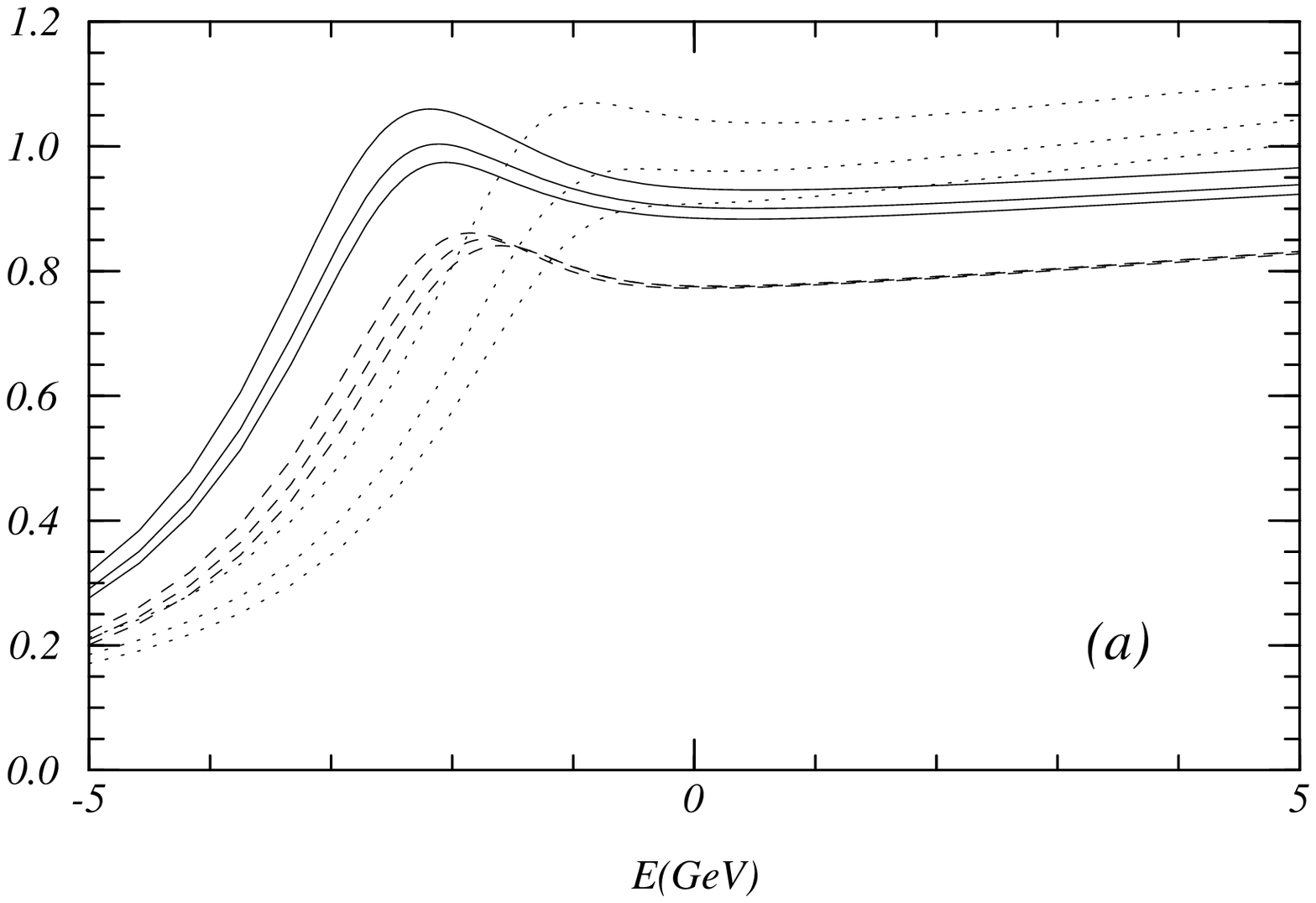}\\
\vspace{3.5cm}
\leavevmode
\epsfxsize=3.5cm
\epsffile[220 420 420 550]{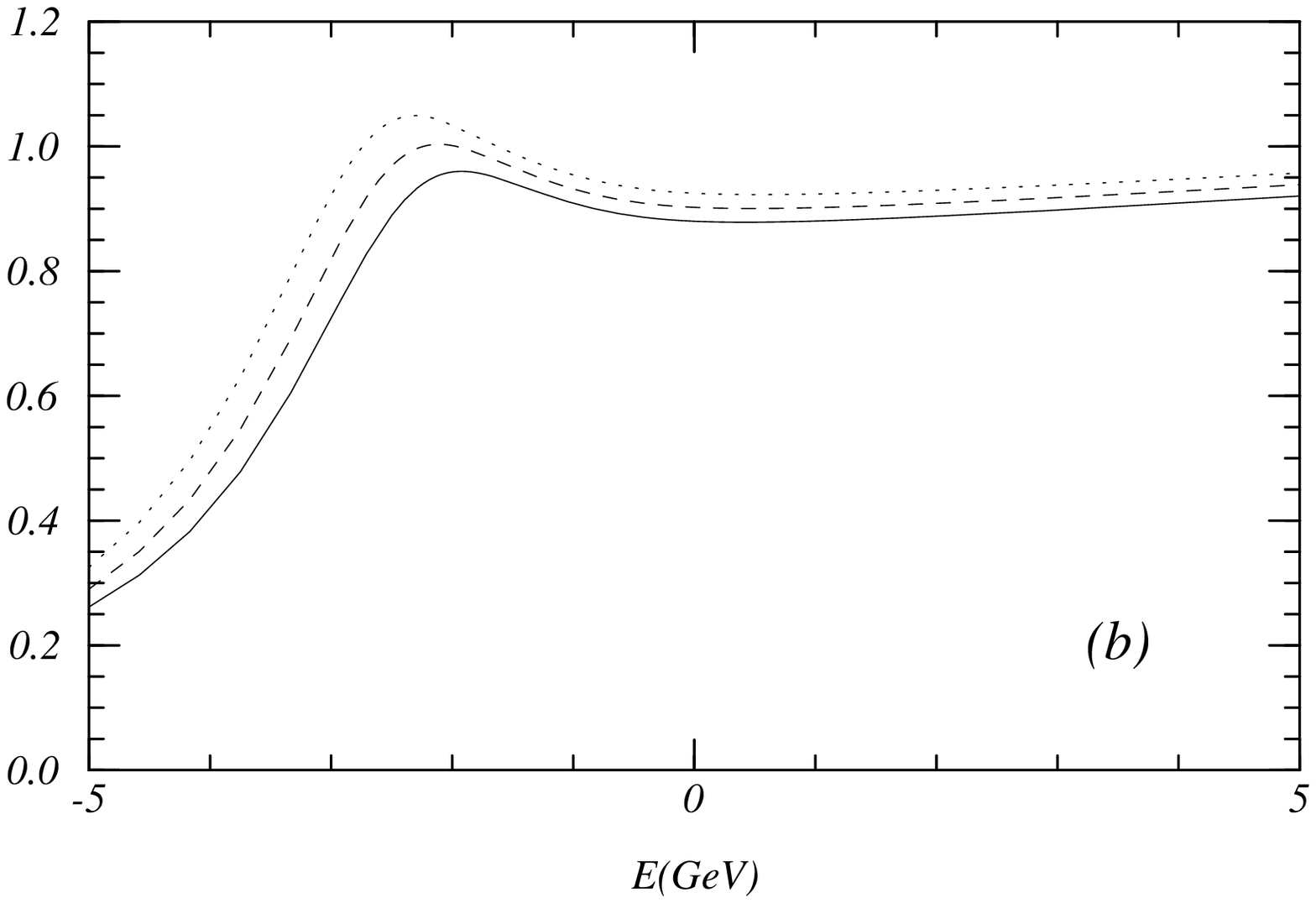}
\vskip 3.5cm
\caption[]{\label{fig9} 
(a) The total normalized photon-mediated $t\bar t$ cross section at
LO (dotted lines), NLO (dashed lines) and NNLO (solid lines)
for the scales $\mu_{\rm soft}=50$ (upper lines), $75$
and $100$~GeV (lower lines). 
(b) The NNLO cross section for $\alpha_s(M_Z)=0.115$ (solid line),
$0.118$ (dashed line) and $0.121$ (dotted line). ($m_t = 175$ GeV,
$\Gamma_t = 1.43$ GeV.  Figures taken from \cite{HT2}.)}
\end{center}
\end{figure}

\noindent
{\bf Calculation and results.}
The problem can be formulated most transparently in the framework of
effective field theories.  There one makes use of the strong hierarchy
of the physical scales top mass, momentum, kinetic energy and 
$\Lambda_{\rm QCD}$ with $m_t \gg m_t v \gg m_t v^2 \gg \Lambda_{\rm
  QCD}$ by integrating out ``hard'' gluons with momenta large compared
to the scales relevant for the nonrelativistic $t \bar t$ dynamics.
This leads to non-relativistic QCD (NRQCD) \cite{CLandBBL}.  With $m_t
v \gg \Lambda_{\rm QCD}$ one can go one step further and integrate out
gluonic (and light quark) momenta of order $m_t v$.  Doing so one
arrives at the so-called potential NRQCD \cite{PinedaSoto}, and the
dynamics of the $t \bar t$ system can be described by the NNLO
Schr\"odinger equation
\begin{eqnarray}
\left[ -\frac{\vec\nabla^2}{m_t} - \frac{\vec\nabla^4}{4m_t^3} +
V_{\rm C}(\vec r\,) + V_{\rm BF}(\vec r\,) + V_{\rm NA}(\vec r\,)  
-\left(E+i\Gamma_t\right)\,\right]\,G(\vec r,E + i \Gamma_t)
\, & = & \, \nonumber\\
= \delta^{(3)}(\vec r\,)\,. \qquad & & 
\label{schroedingerfull}
\end{eqnarray}
Note the appearance of the operator $-\vec\nabla^4/(4m_t^3)$ which is
a correction to the kinetic energy.  The instantaneous potentials are
the two-loop corrected Coulomb potential $V_{\rm C}$
\cite{PeterSchroeder}, the Breit-Fermi potential $V_{\rm BF}$ known
from positronium, and $V_{\rm NA}$ is an additional purely non-Abelian
potential.  The cross section is again related to the imaginary part
of the Green function at $\vec r = 0$.  In contrast to the NLO
calculation the additional potentials lead to ultraviolet divergencies in
Eq.~(\ref{schroedingerfull}) which have to be regularized.  This can
be done by introducing a factorization scale $\mu_{\rm fac}$ which
serves as a cut-off in the effective field theory.  The complete
renormalization also requires the matching of the effective field
theory to full QCD.  This involves the determination of (energy
independent) short distance coefficients.  They contain all
information from the ``hard'' momenta integrated out before and also
depend on the cut-off $\mu_{\rm fac}$, so that in the final result the
biggest part of the factorization scale dependence cancels.  In
order to perform this matching the knowledge of the corresponding NNLO
results of the $t\bar t$ cross section in full QCD above threshold is
essential \cite{CM2}.  Let me skip further details and immediately
discuss the results of the NNLO calculation\footnote{A more detailed
  discussion and complete formulae can be found in \cite{HT2} (see also
\cite{Andre}).}:  Fig.~\ref{fig9}a shows the total cross section $e^+ e^- \to
\gamma^* \to t \bar t$ in units of $\sigma_{\rm point} = 4\pi\alpha^2/(3s)$ 
in LO, NLO and NNLO (dotted, dashed and solid lines, respectively),
where in each case the three curves correspond to three values of the 
scale $\mu_{\rm soft}$ governing the strong coupling in the
potential(s).  In Fig.~\ref{fig9}b the dependence of the NNLO
prediction on the input parameter $\alpha_s(M_Z)$ is demonstrated.  
These results are somewhat surprising:  whereas large corrections are
not unusual for NLO calculations, the large corrections arising at NNLO were
unexpected.  It is well visible from Fig.~\ref{fig9}a that from
leading to NLO the $1S$ peak is shifted to lower energies by about 1
GeV and again moves by about 300 MeV if one includes the NNLO
corrections.  Moreover, the large negative correction in the
normalization from leading to
NLO is partly compensated by the big positive correction at NNLO.  In
addition the scale uncertainty, which is often used as an estimate of
the uncertainty of a (fixed order) perturbative calculation from
higher orders, seems to be artificially small at NLO but fairly big
again at NNLO.  This will make studies which mainly depend on
the normalization of the $t\bar t$ cross section (like the extraction
of the Higgs mass) very difficult.
\begin{figure}[htb]
\begin{center}
\leavevmode
\epsfxsize=8.cm
\epsffile[72 235 540 555]{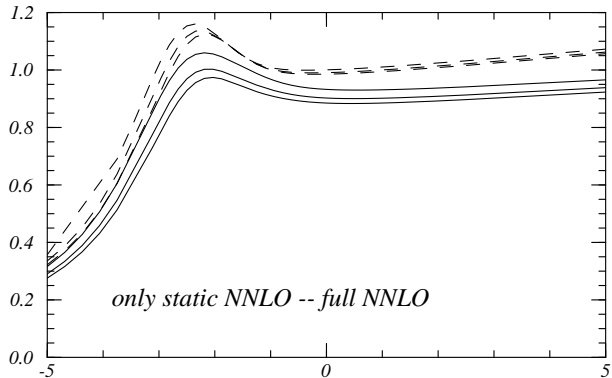}
\caption[]{\label{fig10} Total cross section at NNLO as a function of
  the energy relative to threshold with parameters as in Fig.~\ref{fig9}a.
  The solid lines give the complete result of \cite{HT2} whereas the
  dashed lines contain only the NNLO corrections to the static Coulomb
  potential $V_{\rm C}$ \cite{PeterSchroeder}.}
\end{center}
\end{figure}

In Fig.~\ref{fig10} the importance of the NNLO relativistic
corrections to the kinetic energy and through the additional
potentials $V_{\rm BF}$ and $V_{\rm NA}$ in
Eq.~(\ref{schroedingerfull}) are demonstrated:  the dashed lines show
the result where only the NNLO corrections to the static Coulomb
potential $V_{\rm C}$ \cite{PeterSchroeder} are applied, the solid
lines show the complete NNLO result from \cite{HT2}.

We have argued above that the total cross section with
its steep rise in the threshold region (the remainder of the $1S$ peak
as shown in Fig.~\ref{fig2}) is the ``cleanest'' observable to
determine $m_t$.  From Fig.~\ref{fig9} it now becomes clear that the
problem of the strong correlation between $m_t$ and $\alpha_s$, which
was already discussed above, also appears through the different orders
of perturbation theory:  a fit of experimental data from a threshold
scan to theoretical predictions (like indicated in Fig.~\ref{fig2}) at
a given order will result
in a determination of $m_t$ depending on the order.  This is in principle
nothing wrong and is easily understood, as in higher orders the corrections
to the potential lead to a stronger effective coupling.  Nevertheless
now the question arises:\\ 

\noindent
{\bf Are there large theoretical uncertainties in the determination 
of $m_t$?}  First I would like to point out that the $1S$ peak shift
from NLO to NNLO is actually not too dramatic.  Taking this shift as an
estimate of unknown effects in even higher orders would indicate a
theoretical uncertainty 
$\Delta m_t \stackrel{\scriptstyle <}{\scriptstyle \sim} \Lambda_{\rm
  QCD}$, which still leads to a relative accuracy of $\Delta m_t /m_t
\sim {\cal O}(10^{-3})$ for the top mass.  Still, having argued that
due to the large
width $\Gamma_t > \Lambda_{\rm QCD}$ non-perturbative effects should
be suppressed, an even smaller theoretical uncertainty should be
achievable.  Concerning the large NNLO corrections to the
normalization and the large scale uncertainty I would like to
comment that there is reason to believe that the NNLO result is a much
better approximation than the NLO one and that corrections in even
higher orders should not spoil this picture \cite{HT3}.  But how can
the stability of the prediction be improved?  The key point here is to
remember that in all formulae and results discussed up to now $m_t$ is
defined as the pole mass.  This scheme seems, at first glance, to
be the most intuitive one and to be suited for the non-relativistic
regime.  Nevertheless we know that $m^{\rm pole}$ is {\em not} an
observable.  It is defined only up to uncertainties of 
${\cal O}(\Lambda_{\rm QCD})$, and the large top quark width $\Gamma_t$
does not protect the pole mass $m_t^{\rm pole}$ \cite{SW}.  By
performing a renormalon analysis it was recently shown in \cite{B,
  HSSW} that the leading long-distance behaviour which affects the
pole mass in higher orders also appears in the static potential.
However, in the sum $E_{\rm static} = 2 m^{\rm pole} + E_{\rm binding}$
these contributions cancel and $E_{\rm static}$ is free from 
renormalon ambiguities.  The separate quantities, mass and potential,
suffer from a scheme ambiguity which is not present in the sum.
Therefore one should make use of a ``short distance'' mass definition different
from the pole mass scheme, which avoids these large distance ambiguities.\\

\begin{figure}[htb]
\begin{center}
\leavevmode
\epsfxsize=9.cm
\epsffile[90 95 470 705]{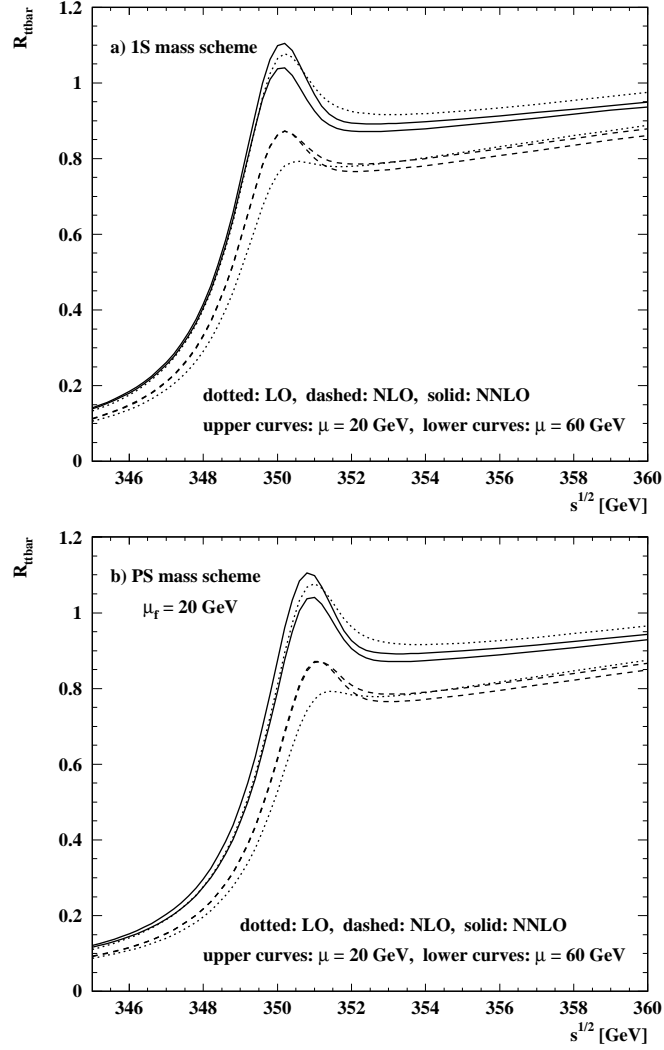}
\caption[]{\label{fig11} Total cross section $e^+ e^- \to
\gamma^* \to t \bar t$ in units of $\sigma_{\rm point} =
4\pi\alpha^2/(3s)$ as a function of $\sqrt{s}$.  Dotted, dashed and solid
lines correspond to the LO, NLO and NNLO results.  The upper curves
are obtained with the renormalization scale $\mu = 20$ GeV, the lower
ones with $\mu = 60$ GeV.  a) $1S$ mass scheme, and b) PS mass scheme
with $\mu_f = 20$ GeV.  ($m_t = 175$ GeV, $\Gamma_t = 1.43$ GeV and
$\alpha_s(M_Z) = 0.118$.)}
\end{center}
\end{figure}

\noindent
{\bf Short distance mass definitions: Curing the problem.}  In
principle there exist infinitely many mass definitions which subtract
the renormalon ambiguities.  In practice, however, this is not
enough.  On the one hand, any new short distance mass $m^{\rm SD}$ has to
be related with high accuracy to a mass in a more general scheme like
the (modified) Minimal Subtraction scheme ($\overline{\rm
  MS}$).\footnote{This is possible because of the short distance
  characteristics of $m^{\rm SD}$ and $m^{\overline{\rm MS}}$ which makes the
  perturbative relation between the masses well behaved.  The
  $\overline{\rm MS}$ mass itself cannot be used directly for
  the calculation of the $t\bar t$ threshold, see \cite{B}.}
Otherwise the extraction of $m^{\rm SD}$ would be more or less useless.
On the other hand, the subtraction of renormalon contributions, which
become important at high orders of perturbation theory, will not be
enough to compensate the large shifts of the $1S$ peak observed at 
NLO and NNLO.  Recently different mass definitions were proposed
which can fulfill all the requirements:  in Ref.~\cite{B} Beneke
defined the ``Potential Subtracted'' mass by
\begin{equation}
m^{\rm PS}(\mu_f) = m^{\rm pole} - \delta m(\mu_f)
\end{equation}
where the subtraction is given by
\begin{equation}
\delta m(\mu_f) = -\frac{1}{2}\,\int_{|\vec q\,|<\mu_f} \frac{{\rm d}^3
  q}{(2\pi)^3}\,\tilde V(q)\,.
\end{equation}
The subtracted potential in position space then reads
\begin{equation}
V(r,\,\mu_f) = V(r) + 2 \delta m(\mu_f)\,.
\end{equation}
This is equivalent to suppressing contributions from momenta $q$ below
the scale $\mu_f$ in the potential.  For $\mu_f \to 0$ one recovers
the pole mass $m^{\rm PS} \to m^{\rm pole}$.  By choosing $\mu_f$
larger, say 20 GeV, one can achieve a compensation of the $1S$ peak
shifts.  Another mass definition is the
$1S$ mass, originally introduced in $B$ meson physics \cite{HLM},
which defines the $1S$ mass as half of the perturbatively defined $1S$
energy.  This $m^{1S}$ mass can be related reliably to the $\overline{\rm
  MS}$ mass.  There are also other mass definition in the literature,
see e.g.\ the ``low scale running mass'' \cite{Uraltsevetal}, which is 
similar to the concept of the PS mass but differs in the actual
$\mu_f$-dependent subtraction.  Studies about the application of
different mass definitions are underway and I can only present
preliminary results here:  Fig.~\ref{fig11} shows our best prediction
\cite{HT3} for the NNLO $t\bar t$ cross section together with the NLO
and LO results for two different values of the renormalization scale
$\mu_{\rm soft}$ governing the strong coupling $\alpha_s$.
In the upper plot the $1S$ mass scheme is used, whereas for the lower
plot the PS mass scheme is adopted.  It is clear from these curves
that both mass definitions work well.  The shift of the $1S$ peak
is nearly completely compensated.  Differences in the normalization
remain, but they will not spoil the mass determination from the
shape of the total cross section near threshold.  Of course more detailed
studies are needed to find the best strategy for a precise
determination of $m_t^{\overline{\rm MS}}$, which is needed in
electroweak calculations.

\section{Studies above Threshold}
In the continuum top quarks are produced through the same annihilation
process as near threshold:  $e^+ e^- \to \gamma^*,\, Z^* \to t \bar
t\,$.  Other (gauge boson fusion) channels like $\ e^+ e^- \to \nu_e
\bar\nu_e t\bar t\ $ or $\ e^+ e^- \to e^+\bar\nu_e t\bar b\ $ are
negligible, except for $e^+ e^- \to e^+ e^- t\bar t\,$, where the
contribution from $\gamma\gamma$ fusion becomes important at TeV
energies.  Formulae for the (polarized) production cross section and
subsequent decay are well known (see e.g.~\cite{report123A} and
references therein).  Similar to the top quark analyses at Fermilab
$t\bar t$ events will be reconstructed at an event by event basis and
allow for a determination
of the top quark mass and its couplings.  Due to the clean environment
and the large statistics (at $\sqrt{s} = 500$ GeV and with an integrated
luminosity of $\int{\cal L} = 50\ {\rm fb}^{-1}$ there will be 
$\stackrel{\scriptstyle >}{\scriptstyle \sim} 30000\ t\bar t$ pairs!)
high precision will be reached at a
future Linear Collider.  In the following I will briefly outline a few
important cases of top physics above threshold.\\

\noindent
$\bullet$ {\bf Kinematical reconstruction of $m_t$ above threshold.}
The top can be reconstructed from 6 jet and 4jet+$l$+$\nu$ events.  For
centre of mass energies far above threshold the top and antitop
signals will be in different hemispheres and $t$ and $\bar t$ may be
reconstructed separately.  Constraints from energy and momentum
conservation in the fitting procedure  can improve the mass resolution
considerably. Experimental studies \cite{report123A} (see also
\cite{report123E}) have
demonstrated that a high statistical accuracy of the order of $\Delta
m_t ({\rm stat.}) \sim 150$ MeV can be achieved at a future Linear
Collider.  But in contrast to the analysis at threshold many
experimental uncertainties and not very well known hadronization
effects will limit the total expected accuracy to $\Delta m_t \sim
0.5$ GeV.\\

\noindent
$\bullet$ {\bf Top formfactors.}  Top quarks are produced with a high
longitudinal polarization.  Due to the large top width $\Gamma_t$
hadronization is suppressed and the initial helicity is transmitted to
the final state without depolarization.  Therefore, in contrast to the
case of light quarks, $t$ helicities can be determined from the
(energy-angular) distributions of jets and leptons in the decay $t \to
b W^+ \to b f \bar f'\,$, similar to the case of $Z$ polarization
analyses at LEP and SLC.  This will allow to measure the formfactors
of the top quark in detail \cite{report123D}.  The relevant current
can be written as
\begin{equation}
j_{\mu}^a \propto \gamma_{\mu} \left( F_{1, L}^a P_L + F_{1, R}^a P_R
  \right) + \frac{i \sigma_{\mu\nu} q^{\nu}}{2m_t} \left( F_{2, L}^a
  P_L + F_{2, R}^a P_R \right)\,,
\end{equation}
with the form factors $F^a$ ($a = \gamma,\,Z,\,W$).  At lowest order
in the SM, $F_{1, L}^{\gamma} = F_{1, R}^{\gamma} = F_{1, L}^W = 1\,$, 
$F_{2, L}^{\gamma} = F_{2, R}^{\gamma} = F_{1, R}^W = 0\,$ and 
$F_{1, L}^Z = g_L$, $F_{1, R}^Z = g_R$.  A non-zero value for 
$(F_{2, L}^{\gamma,Z} + F_{2, R}^{\gamma,Z})$ is caused by a
magnetic ($\gamma$) or weak ($Z$) dipole moment, whereas a non-zero
value for the CP-violating combination 
$(F_{2, L}^{\gamma,Z} - F_{2, R}^{\gamma,Z})$ by an electric (weak)
dipole moment.  These moments would influence distributions for the
top production process, e.g.\ by inducing an extra contribution
proportional to $\sin^2\theta$ in the differential cross section:
\begin{equation}
\frac{{\rm d}\sigma}{{\rm d}\cos\theta} \propto \left[ \frac{m_t}{E}
  \left( F_{1,L}+F_{1,R} \right) + \frac{E}{m_t} 2 \left( F_{2,L}
  +F_{2,R} \right) \right]^2 \sin^2\theta\,.
\end{equation}
The extra ($F_{2,L} +F_{2,R}$) term leads to an additional spin-flip
contribution and therefore changes the total and differential cross
section. At a future Linear Collider such an anomalous magnetic
moment of the top quark $(g-2)_t$ could be seen up to a limit of 
$\Delta\delta 
\stackrel{\scriptstyle <}{\scriptstyle \sim}4\%$ ($\delta \equiv 
F_{2,L}^{\gamma} + F_{2,R}^{\gamma}$) \cite{report123D},
for $\int{\cal L} = 50$ fb$^{-1}$ at $\sqrt{s} = 500$ GeV.  
With especially defined observables an anomalous electric and weak
dipole moment due to CP violating
formfactors $\delta_t^{\gamma, Z} \propto 
(F_{2, L}^{\gamma,Z} - F_{2, R}^{\gamma,Z})$ could be
observed up to a limit of $\Delta d_t^{\gamma, Z} 
\stackrel{\scriptstyle <}{\scriptstyle \sim} 5 \cdot 10^{-18} {\rm e
  cm}$ (for $\int{\cal L} = 10$ fb$^{-1}$ at $\sqrt{s} = 500$ GeV).

A measurement of $F_{1,R}^W \neq 0$ would signal non-SM physics like a
($V$+$A$) admixture to the top charged
current, a $W_R$ boson or the existence of a charged Higgs boson.
$F_{1,R}^W$ can be studied by help of the energy and angular
distributions of the top quark decay leptons \cite{JK3}.  
It could be constrained up to $\Delta \kappa^2 
\stackrel{\scriptstyle <}{\scriptstyle \sim} 0.02$ ($\kappa^2 \sim
|F_{1,R}^W|^2$) with a luminosity of $\int{\cal L} = 50$ fb$^{-1}$ in
the threshold regime, which is best suited for such a measurement.\\

\noindent
$\bullet$ {\bf Rare top decays.}
In the SM top quark decays different from $t \to b W^+$ are strongly
suppressed.  On one hand, the unitarity of the CKM matrix constrains
$V_{tb} \simeq 0.999$, giving not enough room for top decays to the
$s$ or $d$ quark at an observable rate.  On the other hand, due to the GIM
mechanism \cite{GIM}, flavour-changing one-loop transitions like $t
\to c g$, $t \to c \gamma$, $t \to c Z$ or $t \to c H$ are also
extremely small \cite{EHS, MPS}.  However, in extensions of the SM like the 
MSSM extra top quark decay channels like $t \to b H^+$, $t \to
\tilde t \tilde\chi^0,\ \tilde b \tilde \chi_1^+$ may be open.
In general branching fractions of up to 30\% are possible.  The
experimental signatures are clear and will be easily detectable
\cite{VB, PRep}. With an integrated
luminosity of $\int{\cal L} = 50$ fb$^{-1}$ it will be possible to
observe $t \to b H^+$ up to $m_{H^+} 
\stackrel{\scriptstyle <}{\scriptstyle \sim} m_t - 15$ GeV, and $t \to
\tilde t \tilde \chi^0$ down to a branching fraction of $\sim 1\%$ 
at the 3$\sigma$ level.\\

\noindent
$\bullet$ {\bf Direct observation of the top Yukawa coupling.}
Although the Higgs boson will hopefully be discovered before the
future Linear Collider starts operation, the detailed study of the
Higgs and its couplings will remain one of the main tasks of the LC.
There one will be able to test if the Higgs Yukawa coupling to the top quark
deviates from the SM value $\lambda_t^2 = \sqrt{2}\,G_F\,m_t^2
\sim 0.5$.  Studies at threshold will be difficult (see above), but 
due to this large coupling (in comparison to 
$\lambda_b^2 \sim 4\cdot 10^{-4}$) the $t\bar t H^0$ vertex will be
accessible through Higgs-strahlung at high energies.  For $M_H \leq
2m_t$ one will measure $\lambda_t^2$ through the process
$e^+ e^- \to t\bar t H$ with the Higgs subsequently decaying into a
pair of $b$ quarks.  For $M_H \geq 2m_t$ two different processes will
be dominant:  Higgs radiation from $Z$ (in $e^+ e^- \to Z H$) with
subsequent decay of the Higgs into $t\bar t$, and the fusion of $W^+
W^-$ (in $e^+ e^- \to \nu \bar\nu H$) into the Higgs which then decays
into $t\bar t$.  With eight jets in the final state of the fully
hadronic decay channels, which satisfy many constraints, these
processes will have clear signatures.  Still, even despite the large
Yukawa coupling, the cross sections are quite small, amounting only to
a few fb.  Here the planned high luminosity of the latest TESLA design
will be most welcome.  Extensive studies were performed and come to
the conclusion that at high energy and with high luminosity 
$\lambda_t^2$ may finally be measurable with an accuracy of 5\% at a 
future LC \cite{ttbarhiggs}.\\

\section{Conclusions}
I have reviewed the subject of top quark physics at a future $e^+ e^-$
Linear Collider, emphasizing top quark physics at threshold.
Threshold studies will determine the SM parameters $m_t$, $\alpha_s$
and $\Gamma_t$ with very high accuracy:  $\Delta m_t / m_t 
\stackrel{\scriptstyle <}{\scriptstyle \sim} 10^{-3}\,$, 
$\,\Delta\alpha_s \stackrel{\scriptstyle <}{\scriptstyle \sim}
0.003\,$ and $\,\Delta\Gamma_t/\Gamma_t 
\stackrel{\scriptstyle <}{\scriptstyle \sim} 0.05\,$ seem to be
possible from experimental point of view.  Recent theoretical progress
shows, that in order to achieve such a high accuracy also in the
theoretical predictions, mass schemes different from the pole mass should
be employed to disentangle correlations between $m_t$ and $\alpha_s$
as well as infrared ambiguities in the definition of $m_t^{\rm pole}$.

In addition to the total cross section and the momentum distribution
of top quarks also observables like the forward-backward asymmetry, 
polarization and axial contributions are calculated.  These
observables will be accessible by help of large statistics due to 
the high luminosity and by the possibility to have polarized $e^+ e^-$
beams.  Above threshold formfactors of the top quark and the top Yukawa
coupling will be measured.  One may study rare top decays and get
sensitive to non-SM physics.

The future Linear Collider will therefore be {\em the} machine to
study top quark physics in detail, to understand the SM better and
eventually to learn more about what comes beyond it.  I hope to have
shown that top quark physics is an interesting field
both for Theory and Experiment.  Further work will be needed to
understand the heaviest known particle better, before data become
available.\\

\noindent
{\bf Acknowledgements}\\[1mm]
It is my great pleasure to thank the organizers for having made the 
{\em Cracow Epiphany Conference '99}\ such an enjoyable and stimulating 
event.  I would also like to express my gratitude to all friends and 
colleagues I have worked with on various topics about toppik 
and top peaks reported here for their fruitful collaboration. 

While writing this contribution I was hit by the shock of the tragic death of 
Bj{\o}rn H.\ Wiik.  His outstanding efforts for the future Linear
Collider and his fascinating personality will be missed.


\begin{thebibliography}{999}

\def\prep#1#2#3{{\it Phys.\ Rep.\ }{\bf #1} (#2) #3}
\def\app#1#2#3{{\it Acta\ Phys.\ Polon.\ }{\bf B #1} (#2) #3}
\def\apa#1#2#3{{\it Acta Physica Austriaca\ }{\bf#1} (#2) #3}
\def\npb#1#2#3{{\it Nucl.\ Phys.\ }{\bf B #1} (#2) #3}
\def\plb#1#2#3{{\it Phys.\ Lett.\ }{\bf B #1} (#2) #3}
\def\prd#1#2#3{{\it Phys.\ Rev.\ }{\bf D #1} (#2) #3}
\def\pR#1#2#3{{\it Phys.\ Rev.\ }{\bf #1} (#2) #3}
\def\prl#1#2#3{{\it Phys.\ Rev.\ Lett.\ }{\bf #1} (#2) #3}
\def\sovnp#1#2#3{{\it Sov.\ J.\ Nucl.\ Phys.\ }{\bf #1} (#2) #3}
\def\yadfiz#1#2#3{{\it Yad.\ Fiz.\ }{\bf #1} (#2) #3}
\def\jetp#1#2#3{{\it JETP\ Lett.\ }{\bf #1} (#2) #3}
\def\zpc#1#2#3{{\it Z.\ Phys.\ }{\bf C #1} (#2) #3}

\vspace{0.5cm}

\bibitem{K}
J.H.~K\"uhn, \app{12}{1981}{347}; {\it Act.\ Phys.\ Austr.}
\ Suppl.\ XXIV (1982)
203.\\
I.Y.~Bigi, Yu.L.~Dokshitzer, V.A.~Khoze, J.H.~K\"uhn, and P.M.~Zerwas,
\plb{181}{1986}{157}.

\bibitem{Designrep}
R.~Brinkmann et al. (Eds.), {\em Conceptual Design of a 500 GeV $e^+ e^-$ 
Linear Collider with Integrated X-ray Laser Facility}, DESY Orange
Preprint DESY 1997-048 and 
ECFA 1997-182.

\bibitem{PRep}
E.~Accomando et al., \prep{299}{1998}{1}.

\bibitem{gammagamma}
I.I.~Bigi, F.~Gabbiani and V.A.~Khoze, \npb{406}{1993}{3}.\\
O.J.P.~Eboli, M.C.~Gonzalez-Garcia, F.~Halzen, and S.F.~Novaes, 
\prd{47}{1993}{1889}.\\
J.H.~K\"uhn, E.~Mirkes and J.~Steegborn, \zpc{57}{1993}{615}.

\bibitem{Kslacrep}
J.H.~K\"uhn, University of Karlsruhe Preprint TTP-96-18 and {\tt
  hep-ph/ 9707321}, published in the proceedings of the {\em 23rd SLAC
  Summer Institute, Stanford, CA, 10-21 July 1995}, SLAC Report 494.

\bibitem{Orr}
V.A.~Khoze, W.J.~Stirling and L.H.~Orr, \npb{378}{1992}{413}.\\
L.H.~Orr, T.~Stelzer and W.J.~Stirling, \plb{354}{1995}{442}.\\
C.~Macesanu and L.H. Orr, Rochester U. Preprint, UR-1542 and 
{\tt hep-ph/ 9808403}.

\bibitem{FK}
V.S.~Fadin and V.A.~Khoze, \zpc{46}{1987}{417} [\jetp{46}{1987}{525}];
\yadfiz{48}{1988}{487} [\sovnp{48}{1988}{309}].

\bibitem{CDF}
CDF Collaboration (F.~Abe, et al.), \prl{82}{1999}{271}.

\bibitem{D0}
D0 Collaboration (B.~Abbott, et al.), FERMILAB-PUB-98-261-E and 
{\tt hep-ex/9808029}.

\bibitem{MM}
A.~Juste, M.~Martinez and D.~Schulte, in {\em $e^+ e^-$ Linear Colliders: 
Physics and Detector Studies}, DESY Orange Report 97-123E;\\
P.~Comas, R.~Miquel, M.~Martinez, and S.~Orteu, in {\em $e^+ e^-$ at TeV 
Energies: The Physics Potential}, DESY Orange Report 96-123D;\\
P.~Igo-Kemenes, M.~Martinez, R.~Miquel, and S.~Orteu, in {\em $e^+ e^-$ at 
500 GeV: The Physics Potential}, DESY Orange Report 93-123C.

\bibitem{FMY}
K.~Fujii, T.~Matsui and Y.~Sumino, \prd{50}{1994}{4341}.

\bibitem{Jnpbproc}
M.~Je\.zabek, {\it Nucl.\ Phys.\ Proc.\ Suppl.}\ {\bf B}37 (1994) 197.

\bibitem{JK}
M.~Je\.zabek and J.H.~K\"uhn, \plb{207}{1988}{91}.

\bibitem{DSao}
A.~Denner and T.~Sack, \npb{358}{1991}{46}.\\
R.~Migneron, G.~Eilam, R.R.~Mendel, and A.~Soni, \prl{66}{1991}{3105}.

\bibitem{JK2}
M.~Je\.zabek and J.H.~K\"uhn, \prd{48}{1993}{1910};
\prd{49}{1994}{4970} (E).

\bibitem{CM}
A.~Czarnecki and K.~Melnikov, BNL-HET-98-21 and {\tt hep-ph/9806244}. 

\bibitem{HJK}
R.~Harlander, M.~Je\.zabek and J.H.~K\"uhn, \app{27}{1996}{1781}, and
references therein.

\bibitem{Kwong}
W.~Kwong, \prd{43}{1991}{1488}.

\bibitem{StrasslerPeskin}
M.J.~Strassler and M.E.~Peskin, \prd{43}{1991}{1500}.

\bibitem{Sumino}
Y.~Sumino, K.~Fujii, K.~Hagiwara, H.~Murayama, and C.K.~Ng, 
\prd{47}{1993}{56}.

\bibitem{JKT}
M.~Je\.zabek, J.H.~K\"uhn and T.~Teubner, \zpc{56}{1992}{653}.

\bibitem{FandB}
W.~Fischler, \npb{129}{1977}{157}.\\
A.~Billoire, \plb{92}{1980}{343}.

\bibitem{Barbierietal}
R.~Barbieri, R.~K\"ogerler, Z.~Kunszt, and R.~Gatto, \npb{105}{1976}{125}.

\bibitem{Sumino2}
H.~Murayama and Y.~Sumino, \prd{47}{1993}{82}.

\bibitem{HJKT}
R.~Harlander, M.~Je\.zabek, J.H.~K\"uhn, and T.~Teubner, \plb{346}{1995}{137}.

\bibitem{JT}
M.~Je\.zabek and T.~Teubner, \zpc{59}{1993}{669}.

\bibitem{MK}
W.~M\"odritsch and W.~Kummer, \npb{430}{1994}{3}.

\bibitem{FKM}
V.S.~Fadin,~V.A.~Khoze and A.D.~Martin, \plb{320}{1994}{141};
\prd{49}{1994}{2247}.

\bibitem{MY}
K.~Melnikov and O.~Yakovlev, \plb{324}{1994}{217}; \npb{471}{1996}{90}.

\bibitem{khoze}
V.A.~Khoze, CERN Preprint CERN-TH-98-176 and {\tt hep-ph/9805505}.

\bibitem{HJKP}
R.~Harlander, M.~Je\.zabek, J.H.~K\"uhn, and M.~Peter, \zpc{73}{1997}{477}.

\bibitem{PS}
M.~Peter and Y.~Sumino, \prd{57}{1998}{6912}.

\bibitem{GK}
R.J.~Guth and J.H.~K\"uhn, \npb{368}{1992}{38}.

\bibitem{BHMandBH}
W.~Beenakker and W.~Hollik, \plb{269}{1991}{425}.\\
W.~Beenakker, S.C. van der Marck and  W.~Hollik, \npb{365}{1991}{24}.

\bibitem{HS}
W.~Hollik and C.~Schappacher, University of Karlsruhe Preprint
KA-TP-3-1998 and {\tt hep-ph/9807427}, {\it Nucl.\ Phys.}\ {\bf B} (in
press).

\bibitem{FKK}
V.S.~Fadin, V.A.~Khoze and M.I.~Kotskii, \zpc{64}{1994}{45}.

\bibitem{P}
M.~Peter, \app{27}{1996}{3805}.

\bibitem{SandJS}
Y.~Sumino, \app{28}{1997}{2461}.\\
Y.~Sumino and M.~Je\.zabek, \app{29}{1998}{1443}.

\bibitem{KT}
J.H.~K\"uhn and T.~Teubner, DESY Orange Preprint DESY-99-031 and 
{\tt hep-ph/9903322}, {\it Eur.\ Phys.\ J.} {\bf C} (in press).

\bibitem{PeterSchroeder}
M.~Peter, \npb{501}{1997}{471}; \prl{78}{1997}{602}.\\
Y.~Schr\"oder, \plb{447}{1999}{321}.

\bibitem{JKPST}
M.~Je\.zabek, J.H.~K\"uhn, M.~Peter, Y.~Sumino, and T.~Teubner,
\prd{58}{1998}{14006}.\\
M.~Je\.zabek, M.~Peter and Y.~Sumino, \plb{428}{1998}{352}.

\bibitem{HT2}
A.H.~Hoang and T.~Teubner, \prd{58}{1998}{114023}.

\bibitem{MYel}
K.~Melnikov and A.~Yelkhovskii, \npb{528}{1998}{59}.

\bibitem{Yakovlev}
O.~Yakovlev, University of W\"urzburg Preprint WUE-ITP-98-036 and 
{\tt hep-ph/9808463}.

\bibitem{BSS}
M.~Beneke, A.~Signer and V.A.~Smirnov, CERN Preprint CERN-TH-99-57 and {\tt
  hep-ph/9903260}.

\bibitem{CLandBBL}
W.E.~Caswell and G.E.~Lepage, \plb{167}{1986}{437}.\\
G.T.~Bodwin, E.~Braaten and G.P.~Lepage, \prd{51}{1995}{1125};
\prd{55}{1997}{5853} (Erratum).

\bibitem{PinedaSoto}
A.~Pineda and J.~Soto, {\it Nucl.\ Phys.\ Proc.\ Suppl.} {\bf B}64
(1998) 428.

\bibitem{CM2}
A.~Czarnecki and K.~Melnikov, \prl{80}{1998}{2531}, and references therein.

\bibitem{Andre}
A.H.~Hoang, \prd{59}{1999}{14039}.\\
A.H.~Hoang, University of California, San Diego, Preprint
UCSD-PTH-98-33 and {\tt hep-ph/9809431}.

\bibitem{HT3}
A.H.~Hoang and T.~Teubner, in preparation.

\bibitem{SW}
M.~Beneke and V.M.~Braun, \npb{426}{1994}{301}.\\
I.I.~Bigi, M.A.~Shifman, N.G.~Uraltsev, and A.I.~Vainshtein,
\prd{50}{1994}{2234}.\\
M.~Beneke, \plb{344}{1995}{341}.\\
M.C.~Smith, S.~Willenbrock, \prl{79}{1997}{3825}.

\bibitem{B}
M.~Beneke, \plb{434}{1998}{115}.

\bibitem{HSSW}
A.H.~Hoang, M.C.~Smith, T.~Stelzer, and S.~Willenbrock, University of
California, San Diego, Preprint UCSD-PTH-98-13 and {\tt
hep-ph/9804227}.

\bibitem{HLM}
A.H.~Hoang, Z.~Ligeti and  A.V.~Manohar, \prl{82}{1999}{277};
\prd{59}{1999}{74017}.

\bibitem{Uraltsevetal}
M.B.~Voloshin, \prd{46}{1992}{3062}.\\
I.~Bigi, M.~Shifman, N.~Uraltsev, and A.~Vainshtein,
\prd{56}{1997}{4017}.\\
A.~Czarnecki, K.~Melnikov and N.~Uraltsev, \prl{80}{1998}{3189}.

\bibitem{report123A}
P.M.~Zerwas (Ed.), {\em $e^+ e^-$ Collisions at 500 GeV: The Physics
  Potential}, Part A, DESY Orange Report DESY92-123A.

\bibitem{report123E}
E.~Accomando, A.~Ballestrero and M.~Pizzio, in {\em $e^+ e^-$ Linear
  Colliders: Physics and Detector Studies}, DESY Orange Report 97-123E.

\bibitem{report123D}
M.~Schmitt, in {\em $e^+ e^-$ at TeV Energies:  The Physics
  Potential}, DESY Orange Report 96-123D.

\bibitem{JK3}
M.~Je\.zabek and J.H.~K\"uhn, \plb{329}{1994}{317};
\npb{320}{1989}{20}.\\
M.~Je\.zabek, \app{26}{1995}{789}.

\bibitem{GIM}
S.L.~Glashow, J.~Iliopoulos and L.~Maiani, \prd{2}{1970}{1285}.

\bibitem{EHS}
G.~Eilam, B.~Haeri and A.~Soni, \prd{41}{1990}{875}.

\bibitem{MPS}
B.~Mele, S.~Petrarca and A.~Soddu, \plb{435}{1998}{401}.

\bibitem{VB}
A.~Venturi, in {\em $e^+ e^-$ at 500 GeV:  The Physics
  Potential}, DESY Orange Report 93-123C.\\
B.~Bagliesi et al., in Ref.~\cite{report123A}.

\bibitem{ttbarhiggs}
T.~Matsui, contribution to these proceedings.\\
M.~Martinez and S.~Orteu, in {\em $e^+ e^-$ at TeV Energies:  The Physics
  Potential}, DESY Orange Report 96-123D.\\
A.~Juste and G.~Merino; M.~Sachwitz, S.~Shichanin and H.J.~Schreiber,
talks presented at the {\em 2nd ECFA/DESY
  Study for Physics and Detectors for a Linear Collider}, 7-10
November 1998, Frascati, Italy. To appear in the proceedings.

\end{thebibliography}
\end{document}